# Lateral Expansion of the Bridges of Cygnus A and other Powerful Radio Sources

By Greg F. Wellman & Ruth A. Daly

Department of Physics, Princeton University, Princeton, NJ 08544

astro-ph/9507104  25 Jul 1995

Measurements of the width of the radio bridge at several locations along the bridge for each of four powerful extended radio sources are presented. Adopting a few simple assumptions, these measurements may be used to predict the radio surface brightness as a function of position across the radio bridge. The predicted and observed surface brightnesses across the bridges are compared and found to agree fairly well. The results are consistent with a simple picture in which the radio power and size of the radio lobe at the forward edge of the radio bridge are roughly time-independent for a given source, and the expansion of the bridge in the lateral direction is adiabatic. There is no indication that reacceleration or energy transport is important in the bridges of these sources. The rate of lateral expansion of the bridge just behind the radio lobe and hotspot in terms of the rate of forward propagation is compared with that predicted, and found to be in good agreement with the predicted value.

## 1. Introduction

The existence of a cocoon inflated by shocked jet material is implied by most models of FRII radio sources. In the standard picture, shocked jet material in the central region of the cocoon is surrounded by a layer of shocked ambient gas. The cocoon is overpressured with respect to the ambient, unshocked, medium.

An overpressured region must expand. References [1] and [2] discuss rough forms for the expansion. Synchrotron emission is thought to be produced by the shocked jet material so lateral expansion of the radio bridge is expansion of the inner cocoon and does not measure the expansion of the outer cocoon. The outer cocoon boundary may be indicated by a discontinuity in the rotation measure as suggested for the bow shock region of Cygnus A [3]. Values of the Hubble constant of 100 km s$^{-1}$ Mpc$^{-1}$ and the deceleration parameter $q_o = 0$ are assumed throughout.

## 2. The Inner Cocoon

The lobe/hotspot region of the inner cocoon is observed to have a complex structure; for example, primary and secondary hotspots, collimated flows and turbulence are observed in various sources, such as Cygnus A [4]. Nonetheless, it is assumed here that the bridge material is quiescent and cylindrically symmetric a suitable distance behind the hotspot. It is also assumed that the magnetic field is tangled and that any lateral slice has constant volume emissivity. Given these assumptions, the surface brightness $S$ in the quiescent part of the bridge is

$$S(x,y) = S(x,0)\sqrt{1 - \left(\frac{2y}{d(x)}\right)^2} \qquad (2.1)$$

where $x$ is distance from the hotspot along the hotspot - core axis, $y$ is the projected height from that axis and $d(x)$ is the bridge diameter. With more assumptions, $S(x,0)$ and $d(x)$ will be related through cooling from adiabatic expansion in §5.

## 3. The Data

Four sources were chosen from the sample of Leahy, Muxlow and Stephens [5]. Three (3C55, 3C265 and 3C356) were chosen because they have the greatest linear extent and are likely to show the greatest lateral expansion. The other was Cygnus A, chosen because the superior resolution might give a better understanding of the phenomenon and the limitations of the simplifying assumptions. Each source has 2 lobes, so 8 separate lobes are studied here.

Using the published maps [5], the apparent $S_{obs}(x,0)$ and FWHM at several positions along each bridge were measured and the model-based one-dimensional deconvolution was applied, letting $d(x)$ and $S(x,0)$ take the values that would produce the observed FWHM and $S_{obs}(x,0)$ after convolution with a gaussian the size of the observing beam. In general, after the deconvolution, the width is decreased (typically by less than 30%) and the peak is increased (typically by less than 70%) from the values obtained directly from the convolved map.

### 3.1. *Widths*

Figure 1 shows the bridge diameter as a function of separation from the hotspot for all eight bridges. The bridge widths increase as the distance from the hotspot increases; this is interpreted as expansion of the bridge in the lateral direction. In general, the expansion is faster near the hotspot and slower far from the hotspot. This agrees with [1] and [2] which both derive relations roughly of the form $v_{exp} \propto t^{-1/2}$. The ratio of lateral expansion velocity to lobe advance velocity is given by half of the slope of the line in figure 1. The initial (i.e. just behind the hotspot) rate of expansion is roughly $v_{exp} = (0.3 \pm 0.1)v_L$ which agrees well with the theoretical prediction of Daly [2].

Two sources deviate from the expectations of the simple model outlined above: (1) The expansion velocity of the southern lobe of 3C265 appears to increase rather than decrease with increasing distance from the hotspot, as can be seen by the increasing slope of the line in figure 1. From an examination of the map, it seems likely that a slight backflow deflected by the galaxy is causing the apparent increase in width; (2) Using the FWHM to estimate the width of the bridge, as described above, Cygnus A (especially the southern lobe) shows a post-hotspot bulge. This likely results from complex structure and large scale flows near the hotspots.

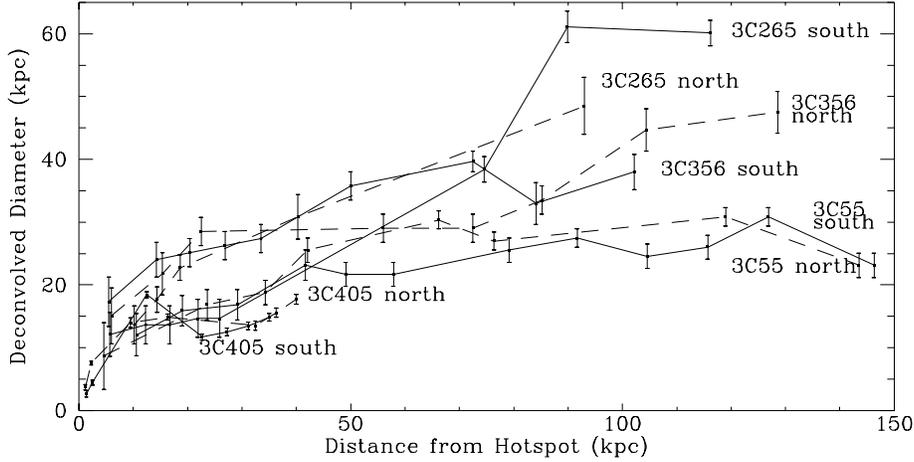

FIGURE 1. Widths as a Function of Distance from Hotspot

## 4. Adiabatic Cooling

Given the assumptions described in §2, and the additional assumptions that the lobe radius and power are time-independent for a given radio source, a relation between the amount of lateral expansion and surface brightness in a quiescent bridge can be derived. The assumption of constant lobe power is consistent with the observed lack of a strong correlation between source length and power [6].

Following [7], consider a volume $V$ containing $N_{tot}$ relativistic electrons with a power law energy distribution $n(\gamma)\,d\gamma = n\gamma^{-s}\,d\gamma$ from $\gamma = \gamma_{co}$ to $\infty$, so

$$N_{tot} = nV \int_{\gamma_{co}}^{\infty} \gamma^{-s}\,d\gamma = nV\gamma_{co}^{-2\alpha}/2\alpha\,, \qquad (4.2)$$

where $\alpha = (s-1)/2$ is the radio spectral index. The radio power from this region is

$$P_\nu = \left[\frac{4\pi e^3}{9mc^2}\right] nV E^{-2\alpha} B_\perp\,, \qquad (4.3)$$

where $E = \gamma m_e c^2$ is the electron energy and the relation between $\nu$ and $E$ is

$$\nu = \frac{3eB_\perp}{4\pi mc}\left(\frac{E}{m_e c^2}\right)^2\,, \qquad (4.4)$$

and $B_\perp$ is the component of the magnetic field perpendicular to the electron motion. For a tangled field, $B_\perp^2 = \frac{2}{3}B^2$.

Following [8,9], the cylindrical expansion of a slice of a bridge with initial diameter $d_0$ to diameter $d$, causes the following changes to occur:

$$\gamma = (d/d_0)^{-2/3}\gamma_0 \; , \tag{4.5}$$

$$B = (d/d_0)^{-4/3}B_0 \; , \tag{4.6}$$

and $V = (d/d_0)^2 V_0$, where the expansion is assumed to be adiabatic; note that a subscript zero refers to quantities prior to expansion and those lacking the subscript zero refer to quantities after the expansion. For constant observing frequency $\nu$, equations (4.4) and (4.6) imply that $E = (d/d_0)^{2/3}E_0$. Equations (4.2) and (4.5) imply that $nV = (d/d_0)^{-4\alpha/3}n_0 V_0$. Substituting these powers of $(d/d_0)$ into equation (4.3) gives the result that:

$$P_\nu = (d/d_0)^{-\frac{8\alpha}{3}-\frac{4}{3}} P_{\nu,0} \; . \tag{4.7}$$

This means that the volume emissivity varies as $\varepsilon_\nu = (d/d_0)^{-\frac{8\alpha}{3}-\frac{10}{3}}\varepsilon_{\nu,0}$, and the surface brightness of a point following the expansion varies as

$$S_\nu = (d/d_0)^{-\frac{8\alpha}{3}-\frac{7}{3}} S_{\nu,0} \; . \tag{4.8}$$

### 4.1. *Comparison of Surface Brightnesses*

An initial surface brightness, $S(x_0, 0)$ can be used to attempt to predict subsequent surface brightness using equation (4.8).

$$S_{predict}(x, 0) = S(x_0, 0) \left(\frac{d(x)}{d(x_0)}\right)^{-\frac{8\alpha}{3}-\frac{7}{3}} \tag{4.9}$$

The idea here is to see whether the decrease of the radio surface brightness across the bridge can be explained by cylindrically symmetric, adiabatic expansion of the bridge. Thus, it is assumed that the surface brightness of the bridge prior to expansion is given by that just behind the hotspot, and the amount of adiabatic expansion is given by the bridge width relative to that just behind the hotspot. Thus, $x_0$ was chosen to be a point one or two contours behind the hot-spot and $S(x_0, 0)$ is taken to be the surface brightness at this location. Of course, any position $x_0$ could have been used to estimate $S(x_0, 0)$. A location just behind the hotspot was chosen to avoid dominance by the hot-spot while testing the model over as much of the bridge as possible. In the absence of precise knowledge of the 178 MHz spectral index across the sources, a constant value of $\alpha = 0.7$ was assumed, which is a typical low-frequency value for these sources.

Figures 2 through 4 compare the observed surface brightness (solid line) with the predicted surface brightness (dashed line) for the eight bridges. The predicted and observed surface brightnesses match each other quite well for several sources. In particular, the four plots from the two sources 3C356 and 3C55 that are of greatest physical length show remarkable agreement. The simplistic model used here does not deal well with the backflow in 3C265 and the post hotspot FWHM bulge in Cygnus A, noted in §3.1. Any form of large scale energy transport or re-acceleration in the cocoon would destroy the relation (4.9). The fact that the comparisons are so close suggests that such processes do not dominate the behavior of the bridges.

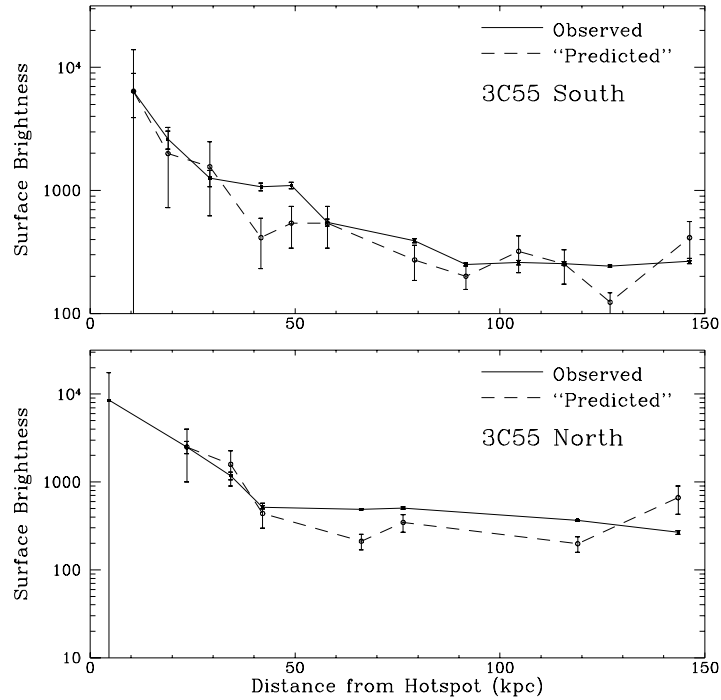

FIGURE 2.

## 5. Conclusions

Models of an overpressured cocoon expanding into the external medium are supported. The ratio of the lateral expansion velocity to the lobe advance velocity can be measured and the initial post-hotspot (i.e. lobe) value agrees well with the predicted value [2]. The bridge surface brightness at position $x$ is predicted by assuming the initial value of the surface brightness was the same as that presently close to the hotspot and that the bridge expanded by an amount given by the ratio of the bridge width at position $x$ relative to that just behind the hotspot. Thus, the model assumes that the lobe radius and radio power are roughly constant over the source lifetime. The remarkably good fit between the model predictions and the data suggest that, for a given radio source, the radio power and lobe radius are time-independent. Significant portions of the bridges of powerful radio sources are "quiescent" and, at low frequencies, adiabatic expansion appears to be the dominant cooling mechanism. Areas close to the hotspots show signs of turbulence or backflow and can not be considered to be quiescent.

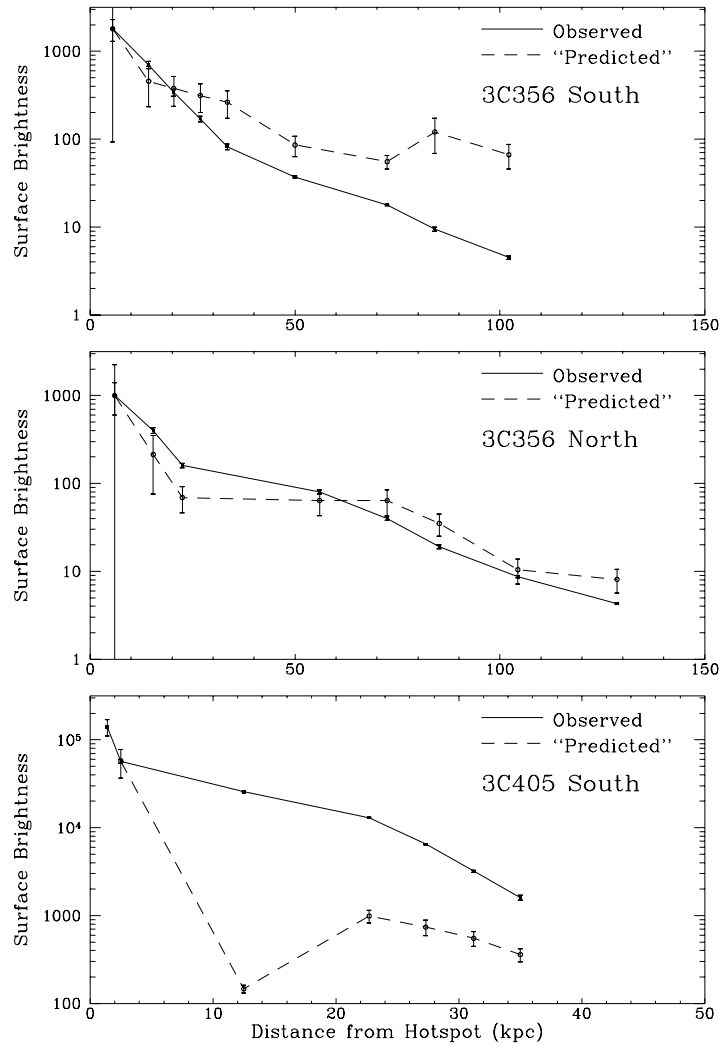

FIGURE 3.

The authors would like to thank Rick Perley, Ed Groth, Lyman Page, Tom Herbig, Lin Wan and Eddie Guerra for helpful discussions. This work was supported in part by the US National Science Foundation, and the Natural Sciences and Engineering Research Council of Canada.

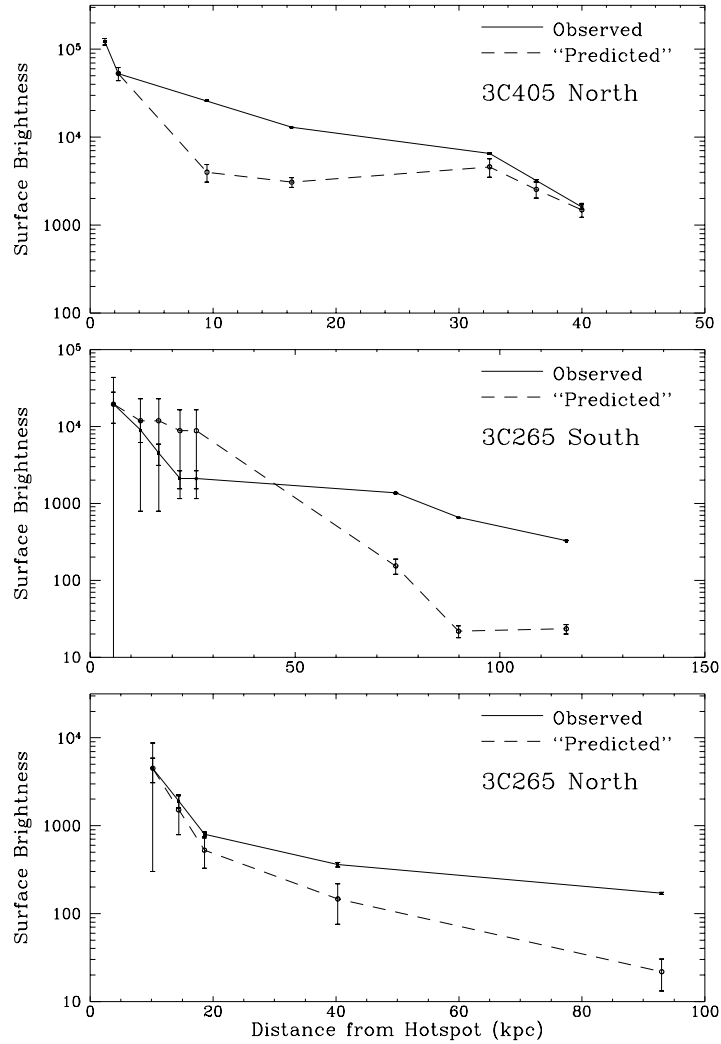

FIGURE 4.